\documentclass[12pt]{article}
\usepackage{epsfig}
\usepackage{a4,isolatin1}
\usepackage{amsmath,amsfonts,latexsym, amssymb}
\newtheorem{satz}{Theorem}[section]

\newtheorem{bem}[satz]{Remark}
\newtheorem{lemma}[satz]{Lemma}

\newtheorem{assumption}[satz]{Assumption}

\newtheorem{conclusion}[satz]{Conclusion}
\newtheorem{ob}[satz]{Observation}

\newcommand{\mcal}{\mathcal}

\newcommand{\tit}{\textit}

\newcommand{\R}{\mathbb{R}}

\begin{document}
\thispagestyle{empty}
\begin{center}
\vspace*{1.0cm}

{\LARGE{\bf Scaling Limit and Renormalisation Group in General
    (Quantum) Many Body Theory}}

\vskip 1.5cm

{\large {\bf Manfred Requardt }}

\vskip 0.5 cm

Institut f\"ur Theoretische Physik \\
Universit\"at G\"ottingen \\
Bunsenstrasse 9 \\
37073 G\"ottingen \quad Germany\\
(E-mail: requardt@theorie.physik.uni-goettingen.de)

\end{center}

\vspace{1 cm}

\begin{abstract}
  Using the machinery of smooth scaling and coarse-graining of
  observables, developed recently in the context of so-called
  fluctuation operators (originally developed by Verbeure et al), we
  extend this approach to a rigorous renormalisation group analysis of
  the critical regime. The approach is completely general,
  encompassing classical, quantum, discrete and continuous systems.
  Our central theme is the analysis of the famous `scaling
  hypothesis', that is, we make a general investigation under what
  cluster conditions of the l-point correlation functions a scale
  invariant (non-trivial) limit theory can be actually attained.

\end{abstract} \newpage

\section{Introduction}One of the crucial ideas of the
renormalization group analysis of, say, the critical regime, is
\tit{scale invariance} of the system in the \tit{scaling limit}. This
is the famous \tit{scaling hypothesis} (as to the underlying working
philosophy compare any good text book of the subject matter like e.g.
\cite{Ma} and references therein). Central in this approach is the
socalled \tit{blockspin transformation \cite{Ka}}. That is,
observables are averaged and appropriately renormalized over blocks of
increasing size. At each intermediate scale a new \tit{effective
  theory} is constructed and the art consists of choosing (or rather:
calculating) the \tit{critical scaling exponents}, so that the
sequence of effective theories converge to a (scale invariant) limit
theory, provided that the start theory lay on the \tit{critical
  submanifold} in the (in general infinite dimensional) parameter
space of theories or Hamiltonians.

Usually the calculations can only be performed in an approximative
way, the main tools being of a perturbative character and are
frequently model dependent. Typically, in the more general discussion,
spin systems are employed, to motivate and explain the calculational
steps. While the general working philosophy, based on the concepts of
\tit{asymptotic scale invariance}, \tit{correlation length} and the
like, is the result of a deep physical analysis of the phenomena,
there is, on the other side, no abundance of both rigorous and model
independent results.

This applies in particular to the control of the convergence of the
scaled $l$-point correlation functions to their respective limits if
we start from a microscopic theory, lying on the \tit{critical
  submanifold}. In this case, correlations are typically long-ranged
and the usual heuristic arguments about the interplay between poor
clustering, on the one side, and formation of \tit{block variables} of
increasing size, on the other side, become rather obscure as one is
usually cavalier as to the interchange of various limit procedures. It
is known, that this is a very dangerous attitude in this regime.

Furthermore, the clustering of the higher correlation functions in the
various channels of phase space may be quite complex and non-uniform
in general.  A concise and selfcontained discussion of the more
general aspects and problems, lurking in the background together with a
series of note and references, can be found in \cite{Pa}, section 7.

Usually, the crucial scaling relation (the \tit{scaling hypothesis})
\begin{equation}W_l^T(Lx_1,\ldots,Lx_l;\mu^*)=L^{-l\cdot n}\cdot
  L^{l\cdot\gamma}\cdot W_l^T(x_1,\ldots,x_l;\mu^*)\end{equation}
which is conjectured to hold at the fixed point (denoted by $\mu^*$ in
the parameter space), is the starting point (or physical input) of the
analysis. Here, $W_l^T$ denote the truncated $l$-point functions (see
below), $L$ is the diameter of the blocks, $Lx_i$ are the respective
centers of the blocks, $n$ is the space dimension, $\gamma$ the
statistical renormalisation exponent. If it is different from $n/2$,
we have an `\tit{anomalous}' scale dimension.

In the following analysis we want to concentrate on the derivation of
such (and similar) scaling relations for the $l$-point functions from
the underlying microscopic characteristics of the theory. We will do
this in a completely general way, that is, the underlying model theory
can be \tit{classical} or \tit{quantum}, \tit{discrete} or
\tit{continuous}. We try to adopt a quite general point of view,
making as few model assumptions as possible. Our strategy is it, to
deal only with the really characteristic (model independent) aspects
of the subject matter.

In this sense our approach is perhaps similar to the one, expounded in
e.g. \cite{Bu} in the context of the analysis of the \tit{ultraviolet
  behavior} in algebraic quantum field theory. A nice review of the
unifying aspects of the cluster of ideas,underlying renormalisation
with applications to numerous fields is also \cite{Lesne}.

At the end of this introduction we want to remark that there exists a
superficially different approach, which is more related to the well
established concepts of renormalisation theory in quantum field theory
(see, for example, \cite{Amit}, \cite{Be} or \cite{Zinn}), that is,
renormalising propagators, Greens's functions and path integrals and
in which scale invariance is present on a more implicit level.

We do not intend (and actually feel unable) to relate all these
different aspects to each other in this paper. We will instead briefly
sketch, what we are going to do in the follow sections. We concentrate
our analysis entirely on the hierarchy of correlation functions which
define the theory. We construct renormalized limit correlation
functions from them which happen to be scale invariant, thus defining
a new limit theory. These limits are set into relation to the degree
of clustering of the original microscopic correlation functions. We do
not openly discuss the flow of, say, the renormalized Hamiltonians
through parameter space. The characteristics of these intermediate
theories are however implicitly given by their hierarchy of
correlation functions as was explained in e.g. \cite{Re1} or
\cite{Bu}.

The same holds for the effective Hamiltonians which emerge on the
intermediate scales in the ordinary approach. In our approach the
effective time evolution is carried over from the microscopic theory
as described in \cite{Re1} or (in a slightly other context) in
\cite{Bu}, see also \cite{Verbeure}. In case we work in an
environment, defined by ordinary Gibbs states, our framework would
exactly yield these effective Hamiltonians. In our paper we treated
the infrared limit. As our scaling approach depends on a continuous
parameter, $R$, which could as well go to zero, a slight extension of
our approach would allow to study also the ultra violet behavior. In
this case we have to work, however, in a space-time picture, that is,
the time coordinates have also to be scaled. The cluster properties
are then expected to be even more intricate as the clustering with
respect to both space and time is of a markedly dynamical nature.
\section{Concepts and Tools}
As to the general framework we refer the reader to \cite{Re1}. One of
our technical tools is a modified (smoothed) version of averaging
(such a possibility is also briefly mentioned in the notes in
\cite{Pa}). Instead of averaging over blocks with a sharp cut off, we
employ a smoothed averaging with smooth, positive functions of the
type
\begin{equation}f_R(x):= f(|x|/R)\quad\text{with}\quad f(s)=
\begin{cases}1 & \text{for $|x|\leq 1$} \\ 0 & \text{for $|x|\geq 2$}
\end{cases}
\end{equation}
(one can choose even more general classes of functions). This class of
scaled functions has a much nicer behavior under Fourier
transformation, as, for example, functions with a sharp cut off, the
main reason being that the tails are now also scaled. We have
\begin{equation}\hat{f}_R(k)=const\cdot R^n\cdot \hat{f}(R\cdot
  k)\end{equation}
Remark: One might perhaps think that this choice of averaging will
lead to a different limit theory. This is however not the
case. Furthermore, the mathematical differences between the two
approaches, that is, using sharp or smooth and scaled cut off
functions, are relatively subtle and not so apparent. We will
investigate these aspects in \cite{Re2}.\vspace{0.3cm}

Another point, worth mentioning, are the implications of translation
invariance. We have for the correlation functions
\begin{equation}W(x_1,\ldots,x_n)=W(x_1-x_2,\ldots,x_{n-1}-x_n)\end{equation}
The truncated correlation functions are defined inductively as follows
(see \cite{Re1})
\begin{equation}W(x_1,\ldots,x_n)=\sum_{part}\prod_{P_i}W^T(x_{i_1},\ldots,x_{i_k})\end{equation}
The (distributional) Fourier transform reads
\begin{equation}\tilde{W}^T(p_1,\ldots,p_l)=\hat{W}^T(p_1,p_1+p_2,\ldots,p_1+\cdots
p_{l-1})\cdot\delta(p_1+\cdots p_l)\end{equation}
The dual sets of variables are
\begin{equation}      y_i:= x_i-x_{i+1}\;,\;q_i=\sum_{j=1}^i p_j\quad i\leq
  (l-1)\end{equation}

\section{The case of Normal Fluctuations}
As in \cite{Re1}, we assume that away from the critical point the
truncated $l$-point functions are integrable, i.e. $\in
L^1(R^{n(l-1)})$, in the difference variables, $y_i:=x_i-x_{i+1}$. As
observables we choose the translates
\begin{equation}A_R(a_1),\ldots,A_R(a_l)\;,\;A_R(a):=R^{-n/2}\cdot\int
A(x+a)f(x/R)d^nx\end{equation}
(where, for convenience, the labels $1\ldots l$ denote also possibly
different observables). We then get (for the calculational details see
\cite{Re1}, the hat denotes Fourier transform, translation invariance
is assumed throughout, the $const$ may change during the calculation
but contains only uninteresting numerical factors):
\begin{multline}\langle A_R(a_1)\cdots A_R(a_l)\rangle^T=const\cdot
  R^{ln/2}\cdot\\\int \hat{f}(Rp_1)\cdots \hat{f}(-R[p_1+\cdots
 + p_{l-1}])\cdot \hat{W}^T(p_1,\ldots,p_{l-1})\cdot
 e^{-i\sum_1^{l-1}p_ia_i}\cdot e^{ia_l\sum_1^{l-1}p_i}\prod dp_i\\
=const\cdot R^{ln/2}\cdot R^{-(l-1)n}\cdot\\\int \hat{f}(p_1')\cdots
\hat{f}(-[p_1'+\cdots+p_{l-1}'])\cdot
\hat{W}^T(p'_1/R,\ldots,p'_{l-1}/R)\cdot e^{-i\sum_1^{l-1}(p'_i/R)a_i}\cdot e^{ia_l\sum_1^{l-1}p'_i/R}\prod dp'_i\end{multline}
We now scale the $a_i$'s like
\begin{equation}a_i:=R\cdot X_i\;,\;X_i\;\text{fixed}\end{equation}
This yields
\begin{multline}\langle A_R(R\cdot X_1)\cdots A_R(R\cdot
  X_l)\rangle^T=\\const\cdot R^{(2-l)n/2}\cdot \int
  e^{-i\sum_1^{l-1}p'_iX_i}\cdot e^{iX_l\sum_1^{l-1}p'_i}\cdot\\
 \hat{f}(p_1')\cdots\hat{f}(-[p_1'+\cdots+p_{l-1}'])\cdot\hat{W}^T(p'_1/R,\ldots,p'_{l-1}/R)\prod dp'_i\end{multline}

As the $\hat{f}$ are of strong decrease and $\hat{W}^T$ continuous and
bounded by assumption, we can perform the limit $R\to\infty$ under the
integral
and get:\\[0.3cm]
Case 1 ($l\ge3$):
\begin{equation}\lim_{R\to\infty} \langle A^F_R(R\cdot X_1)\cdots A^F_R(R\cdot
  X_l)\rangle^T=0\end{equation}
Case 2 ($l=2$):
\begin{equation}\lim_{R\to\infty}\langle A^F_R(R\cdot X_1)A^F_R(R\cdot
  X_2)\rangle^T=const\cdot\int \hat{W}^T(0)\cdot
  e^{-ip'_1(X_1-X_2)}\cdot\hat{f}(p'_1)\cdot\hat{f}(-p'_1)dp'_1\end{equation}
\begin{conclusion}In the normal regime, away from the critical point,
  where we assumed $L^1$-clustering, all the truncated correlation
  functions vanish in the limit $R\to\infty$ apart from the $2$-point
  function. We hence have a quasi free theory in the limit as
  described in \cite{Re1} or in the work of Verbeure et al (cf. the
  references in \cite{Verbeure})
\end{conclusion}
\section{The Relation to the Heuristic Scaling\\ Hypothesis}
In the following sections we will develop two rigorous, technically
slightly different, approaches which implement the physically
well-motivated but, nevertheless, to some extent heuristic scaling
hypothesis. The analysis will be performed in coordinate
space, Fourier space, respectively. In this section we
restrict ourselves to the two-point correlation function, for which the
asymptotic behavior is considerably simpler and more transparent.

The pedagogical reason for representing two technical methods is the
following. It turns out that, as may be expected on physical grounds,
critical systems and their characteristics are sitting exactly on
points in the parameter space where also the necessary mathematical
methods turn out to be of a singular character. This can best be
exhibited by
contrasting the two methods developed below.\\[0.3cm]
Remark: In the rest of the paper we replace the exponent $n/2$ in the
definition of $A_R(a)$ by an exponent $\gamma$, which will usually be
fixed during or at the end of a calculation. It plays the role of a
\tit{critical scaling exponent}.\\[0.3cm]
Let us hence study the behavior of
\begin{multline}\label{two} \langle A_R(R\cdot X_1)A_R(R\cdot
  X_2)\rangle^T=R^{-2\gamma}\cdot\int W^T((x_1-x_2)+R(X_1-X_2))\\\cdot
  f(x_1/R)f(x_2/R)dx_1dx_2\\
=R^{-2\gamma+2n}\int W^T(R[(x_1-x_2)+(X_1-X_2)])\cdot
f(x_1)f(x_2)dx_1dx_2
\end{multline}
We make the physically well-motivated assumption that--in the critical
regime--$W^T$ decays
asymptotically like some inverse power, i.e.
\begin{equation}W^T(x_1-x_2)\sim F(x_1-x_2)\cdot
  |x_1-x_2|^{-(n-\alpha)}\quad 0<\alpha<n\end{equation}
for $|x_1-x_2|\to\infty$, $F$ bounded and well-behaved.

From the last line of $(\ref{two})$ we see that, as $f$ has compact
support, we can replace $W^T$, for $(X_1-X_2)\neq 0$ and $R\to\infty$
by its asymptotic expression and absorb the remaining
$(x_1,x_2)$-integration in an appropriate constant. We hence get for
$R$ large:
\begin{equation}\langle A^F_R(R\cdot X_1)A^F_R(R\cdot
  X_2)\rangle^T\approx const\cdot R^{-2\gamma+2n}\cdot
  R^{-(n-\alpha)}\cdot|X_1-X_2|^{-(n-\alpha)}\end{equation}
Choosing now
\begin{equation}\gamma=(n+\alpha)/2\end{equation}
we get a limiting behavior (for $R\to\infty$) as
\begin{equation}\sim|X_1-X_2|^{-(n-\alpha)}\end{equation}

Central in the renormalisation group idea is that systems on the
\tit{critical surface} (i.e., critical systems) are driven towards a
\tit{fixed point}, representing a completely scale invariant
theory. This idea is usually formulated in an abstract parameter space
of, say, Hamiltonians. In our correlation function approach the fixed
point shows its existence via the scaling properties of the
correlation functions, that is
\begin{equation}W^T_2(L\cdot(X-Y);\mu^*)=L^{-2(n-\gamma)}W_2^T(X-Y;\mu^*)\end{equation}
with $\mu^*$ describing the fixed point in the (usually) infinite
dimensional parameter space. We see from the above that this is
exactly implemented by our limiting correlation functions, as we have
(with the choice $\gamma=(n+\alpha)/2$):
\begin{equation}W_2^T(X-Y;\mu^*)\sim|X-Y|^{-(n-\alpha)}\end{equation}
That is, the above scaling limit leads to a limit (i.e. fixed point)
theory, exactly reproducing the asymptotic behavior of the original
(microscopic) theory. But there is yet another important point worth to be
mentioned.

In the more general situation of $l$-point correlation functions we do
not necessarily have such a simple decay behavior. Furthermore, it
would be desirable to have precise information on thresholds up to
which a certain method leads to rigorous and reliable results. We
therefore describe and illustrate two methods with the help of the
relatively transparent behavior of the $2$-point functions, which will
then be applied to the more complex and general situation of the
higher correlation functions. Using the assumed simple asymptotic
behavior of the $2$-point functions, we proceed now as follows.
\subsection{\label{One}Method One}
We assume the existence of a certain exponent, $\alpha$, so that
($x^2$ denoting the vector-norm squared)
\begin{equation}G(x):=W^T(x)\cdot(1+x^2)^{(n-\alpha)/2}=const+F(x)\end{equation}
with a decaying (non-singular) $F$. Fourier transformation then
yields:
\begin{multline}R^{-2\gamma}\cdot\int
  W^T_2((x_1-x_2)+R(X_1-X_2))f(x_1/R)f(x_2/R)dx_1dx_2\\
=R^{-2\gamma}\cdot\int
G((x_1-x_2)+R(X_1-X_2))\cdot[1+((x_1-x_2)+R(X_1-X_2))^2]^{-(n-\alpha)/2}\cdot\\f(x_1/R)f(x_2/R)dx_1dx_2\\
=R^{-2\gamma}\cdot R^{2n-(n-\alpha)}\cdot\int dp\,\hat{G}(p)\cdot
e^{-iRp(X_1-X_2)}\cdot\\
\left[\int e^{-iRp(x_1-x_2)}(R^{-2}+((x_1-x_2)+(X_1-X_2))^2)^{-(n-\alpha)/2}f(x_1)f(x_2)dx_1dx_2\right]\end{multline}
where we made the substitution $x\to R\cdot x$.

We now assume the support of $f$ to be contained in a sufficiently
small ball around zero (or, alternatively, $(X_1-X_2)$ sufficiently
large so that $(x_1-x_2)+(X_1-X_2)\neq 0$ for $x_i$ in the support of
$f$).  With
\begin{equation}\hat{G}(p)=const\cdot\delta(p)+\hat{F}(p)\end{equation}
the leading part in the scaling limit $R\to\infty$ is the
$\delta$-term. Asymptotically we hence get for $R\to\infty$ (setting
$y:=x_1-x_2\;Y:=X_1-X_2$):
\begin{equation}R^{n+\alpha-2\gamma}\cdot const\cdot\int
  |y+Y|^{-(n-\alpha)}\cdot f\ast f(y)dy\end{equation}
with
\begin{equation}f\ast f(y):=\int f(y+x_2)\cdot f(x_2)dx_2\end{equation}

The reason why the contribution, coming from $\hat{F}(p)$, can be
neglected for $R\to\infty$ is the following: $f$ is assumed to be in
$\mcal{D}$; by assumption the prefactor never vanishes on the support
of $f(x_i)$. Hence the whole integrand in the expression in square
brackets is again in $\mcal{D}$ and therefore its Fourier transform is
in $\mcal{S}$, that is, of rapid decrease. We can therefore perform
the $R$-limit under the integral and get a rapid vanishing of the
corresponding contribution in $R$ for each $p\neq 0$. As the
$\delta$-contribution has been extracted from $\hat{G}(p)$, this
proves the statement.

As $f\ast f$ has again a compact support, we have that, choosing
\begin{equation}\gamma=(n+\alpha)/2\end{equation}
the limit correlation function behaves as $\sim
|X_1-X_2|^{-(n-\alpha)}$ as in the above heuristic analysis.
\subsection{Method Two}
As in the case of normal clustering or (\cite{Re1}, last section), one
can, on the other hand, improve the too weak decay of $W^T(x_1-x_2)$ and transform it into an integrable (i.e. $L^1$-) function. So, with a
similar notation as in the preceding subsection, we choose a suitable
exponent $\alpha$ in
\begin{equation}P_{\alpha}(x_1-x_2):=(1+|x_1-x_2|^2)^{\alpha/2}\end{equation}
so that
\begin{equation}G(y):=W^T(y)\cdot P_{\alpha}^{-1}\,\in\,L^1\quad
  (y:=x_1-x_2) \end{equation}

In contrast to \tit{Method One}, there is of course a whole range of
such possible exponents, $\alpha>\alpha_{inf}$, so that
\begin{equation}\label{inf}G(y)=\begin{cases}\in L^1 &
    \text{for}\;\alpha>\alpha_{inf}\\
\not\in L^1 & \text{for}\;\alpha<\alpha_{inf}\end{cases}\end{equation}
Proceding as in \tit{Method One}, we get
\begin{multline}R^{-2\gamma}\int W^T(y+R\cdot Y)\cdot
  f(x_1/R)f(x_2/R)dx_1dx_2=\\R^{-2\gamma}\cdot R^{n+\alpha}\int
  dp\,\hat{G}(p/R)\cdot e^{-ipY}\cdot\left[\int
  e^{-ipy}(R^{-2}+(y+Y)^2)^{\alpha/2}\cdot f\ast f(y)dy\right]
\end{multline}

Again the obvious strategy seems to be to choose
\begin{equation}\gamma=(n+\alpha)/2\end{equation}
and perform the limit $R\to\infty$. With the same support properties
as above, that is, $y+Y\neq 0$ for $x_1,x_2\in$ support of $f$, the
integrand in square brackets is again infinitely differentiable with
respect to $y$. Hence, its Fourier transform is fast decaying in
$p$.\\[0.3cm]
Remark: Note that for $\alpha/2$ non-integer and without the above
support restriction, there will show up a singularity in sufficiently
high orders of differentiation for vanishing $R^{-2}$. One can however
control these singularities and show that the analysis still
goes through in the case where the support condition does not
hold. One gets however some mild constraint on the admissible $\alpha$'s.
\\[0.3cm]
Therefore we can again apply Lebesgues' theorem of dominated
convergence and perform the $R$-limit under the integral. This yields
the expression
\begin{multline}\hat{G}(0)\cdot\int dpe^{-ipY}\cdot\left[\int e^{-ipy}\cdot
  |y+Y|^{\alpha}\cdot f\ast f(y)dy\right]=\\
const\cdot \hat{G}(0)\cdot\int \delta(y+Y)\cdot|y+Y|^{\alpha}\cdot f\ast
f(y)dy=
const\cdot\hat{G}(0)\cdot 0\end{multline}
(as a result of the above support condition).
\begin{conclusion}With $\alpha$ chosen so that $G(y)\in L^1$ and
  $\gamma=(n+\alpha)/2$, the limit can be carried out under the
  integral and yields the result zero. This shows a fortiori that
  there is no $\alpha_{min}$ with the property that there is a
  non-vanishing limit-two-point function. Put differently, we have an
  $\alpha_{inf}$ but no $\alpha_{min}$ (cf. (\ref{inf})).
\end{conclusion}

So, in contrast to \tit{Method One}, the relevant exponent,
$\alpha_{inf}$, is of such a peculiar nature that we definitely cannot
apply the above method of interchange of taking the limit $R\to\infty$
and integration. But nevertheless, we will show that
\begin{equation}\gamma:=(n+\alpha_{inf})/2\end{equation}
is the correct critical scaling exponent leading to a sensible limit
theory and that this $\alpha_{inf}$ is exactly the $\alpha$, we have
determined in \tit{Method One}.

We have learned above that in order to arrive at a non-zero limit
correlation function, we are definitely forbidden to exploit
Lebesgues' theorem of dominated convergence in the above
expression. The reason for the vanishing of the respective expression
was that with
\begin{equation}\lim_{R\to\infty}\hat{G}(p/R)=\hat{G}(0)\end{equation}
we have to evaluate $\int \hat{g}(p)dp$ with
\begin{equation}\label{g}\hat{g}(p):=\int e^{-ip(y+Y)}|y+Y|^{\alpha}\cdot f\ast
f(y)dy\end{equation}
This integral happens to be zero due to the explicit factor,
$|y+Y|^{\alpha}$ and the assumed support properties.

So, we have to investigate what happens for $\alpha=\alpha_{inf}$. As
we learned above that there is no $\alpha_{min}$, we can conclude
\begin{ob}For $\alpha=\alpha_{inf}$, $G_{\alpha}(y)$ is no longer in
  $L^1$, with
\begin{equation}G_{\alpha}(y):=W^T(y)\cdot(1+y^2)^{-\alpha/2}\end{equation}
\end{ob}
We know that for $\alpha<\alpha_{inf}$ the decay of $G_{\alpha}(y)$ is
so weak that the Fourier transform develops a power law singularity in
$p=0$; that is, we can conclude
\begin{lemma}For $\alpha_{inf}-\alpha:=\varepsilon$, $\hat{G}_{\alpha}$
  has a singularity of the form $|p|^{-\varepsilon}$ near $p=0$.

For $\alpha=\alpha_{inf}$ the singularity is of logarithmic type near
$p=0$.
\end{lemma}
This statement can again be proved by a scaling argument. Let
$G_{\alpha}$ have a non-integrable tail of the form
$|y|^{-(n-\varepsilon)}$. For the (distributional) Fourier transform we
then have
\begin{equation}\hat{G}_{\alpha}(\lambda\cdot p)=const\cdot\int
  e^{i\lambda py}\cdot G_{\alpha}(y)dy=const\cdot\lambda^{-n}\cdot\int
  e^{ipy'}\cdot G_{\alpha}(y'/\lambda)dy'\end{equation}
For $\lambda\to 0$ we can, as above, replace $G_{\alpha}$ by its
asymptotic expression, which goes as $|y|^{-(n-\varepsilon)}$ and
conclude that $\hat{G}_{\alpha}(\lambda p)$ contains a leading singular
contribution $\sim\lambda^{-\varepsilon}$ (modulo logarithmic
terms). We hence see that
\begin{equation}\hat{G}_{\alpha}(p)\sim|p|^{-\varepsilon}\end{equation}
near $p=0$ as a distribution (that is, the above resoning is to be
understood modulo the smearing with appropriate test functions; see
e.g. \cite{Gelfand}).
For $\alpha=\alpha_{inf}$, the singularity must be weaker than any
power, that is, must be of logarithmic type.

By \tit{Method One} we get a limit correlation function which clusters as
$|Y|^{-(n-\alpha)}$. One may wonder where this behavior is hidden if
we use \tit{Method Two}. Taking only the singular term in
$\hat{G}(p/R)$ into account, we have (with
$\gamma:=(\alpha_{inf}+n)/2$)
\begin{equation}\lim_{R\to\infty}R^{-2\gamma}\langle A_R(RX_1)\cdot
  B_R(RX_2)\rangle^T\sim\lim_{R\to\infty}const\cdot\int \ln
  (|p|/R)\cdot \hat{g}(p)dp\end{equation}
and $\hat{g}(p)$ as in equation (\ref{g}). We can again neglect the term
\begin{equation}\ln R\cdot\int \hat{g}(p)dp\end{equation}
as $\int \hat{g}(p)dp=0$.

Assuming at the moment that $\alpha$ were an integer (we will get the
general result by a scaling argument), the prefactor $|y+Y|^{\alpha}$ can
be transformed into corresponding $p$-differentiations of $\hat{f\ast
  f}(p)$, which, by partial integration, can then be shifted to
corresponding differentiations of $\ln(|p|)$. This transformation
yields an expression of the type $|p|^{-\alpha}$ times a smooth and
decaying function. That means, we essentially end up with an expression
like
\begin{equation}\int dp\,|p|^{-\alpha}\cdot e^{-ipY}\cdot\left[\int
      e^{-ipy}f\ast f(y)dy\right]\end{equation}
By the same reasoning as above we conclude that the singularity,
$|p|^{-\alpha}$, goes over, via Fourier transform, into a weak decay
proportional to $|X_1-X_2|^{-(n-\alpha)}$, that is, we arrive at the
same result as in \tit{Method One}, whereas the reasoning is a little
bit more tricky.

For a general non-integer $\alpha$ the argument could be made precise by
analysing the distributional character of an expression like
$r^{\beta}$ with $r:=|x|$ and its Fourier transform. As the analysis
is a little bit tedious, we refer the reader to \cite{Gelfand}. On the
other hand, one can use a scaling argument as above (with
$Y:=\lambda\cdot Y_0\,,\,Y_0\;\text{fixed as}\;\lambda\to\infty$). This
yields an asymptotic behavior of the form
\begin{equation}\lambda^{-(n-\alpha)}\cdot\int
  dp\,\ln(|p|)\cdot\left[\int e^{-i\lambda
      p(y+Y_0)}\cdot|y+Y_0|^{\alpha}\cdot f\ast f(\lambda
    y)dy\right]\end{equation}

The evaluation of the integral for $\lambda\to\infty$ can be done as
follows: As $f\ast f$ has compact support, the volume of the support
of $f\ast f(\lambda y)$ shrinks proportional to $\lambda^{-n}$.
Therefore the expression in square brackets scales as
$\sim\lambda^{-n}$.  On the other hand (due to an `uncertainty
principle' argument), its essential $p$-support increases
proportional to $\lambda^n$. That is, the two effects compensate each
other and we have again a large-$Y$ behavior $\sim|Y|^{-(n-\alpha)}$
as before.

We conclude that both methods lead to the same aymptotic scaling
behavior of the renormalized two-point function.
\section{The General Cluster-Analysis at the Critical Point}
We now study the general situation of the presence of some long-range
correlations in the l-point functions. In contrast to the much simpler
situation prevailing in the case of two-point functions, the
clustering may be quite complicated, in particular, the dependence on
the number, $l$, i.e. the number of observables, occurring in the
expressions, may be non-trivial. Therefore, we have to investigate
these aspects in more detail.

From general principles (see e.g. \cite{Ruelle}) we know that in a
\tit{pure phase} there is always a certain degree of clustering. We
make the slightly stronger assumption that it is in some way of the
kind of an inverse power law at infinity (to be specified below). We
want to study the scaling limit of
\begin{equation}\label{cor}\langle A_R(R\cdot X_1)\cdots A_R(R\cdot
  X_l)\rangle^T\end{equation}
with
\begin{equation}A_R(a):=R^{-\gamma}\cdot\int
A(x+a)f(x/R)d^nx\end{equation}
and an, at the moment, unspecified exponent, $\gamma$.

The above expression can be written as
\begin{equation}\int
  W^T((x_1-x_2)+R(X_1-X_2),\ldots,(x_{l-1}-x_l)+R(X_{l-1}-X_l))\cdot
  \prod_{i=1}^l f(x_i/R)\prod_{i=1}^l dx_i\end{equation}
Fourier transformation yields (with $\hat{W}^T(q_1,\ldots,q_{l-1})$
considered as a distribution on $\mcal{S}(\R^{(l-1)n})$)
\begin{multline}\label{fourcor}const\cdot R^{l(n-\gamma)}\cdot\int
  \hat{W}^T(q_1,\ldots,q_{l-1})\cdot
  e^{-i\sum_{j=1}^{l-1}Rq_jY_j}\cdot\\\left[\int
    e^{-i\sum_1^{l-1}Rp_jx_j}\cdot e^{iRq_{l-1}x_l}\cdot
    \prod_{i=1}^{i=l}
    f(x_i)\prod_{i=1}^{i=l}dx_i\right]\prod_1^{l-1}dp_j\end{multline}
with $Y_j:=X_j-X_{j+1}$ and the wellknown relation between the
$q$-variables and the $p$-variables (see e.g. section 2 or \cite{Re1}). For
calculational or notational convenience we will employ both
sets of variables which are linear combinations of each other.

As $f$ is in $\mcal{D}$ by assumption, the Fourier transform of
\hspace{0.2cm}$\prod f(x_i)$\hspace{0.2cm} is in $\mcal{S}$ and the
function in square brackets is a function of
$(Rp_1,\ldots,Rp_{l-1})\;\text{or}\;(Rq_1,\ldots,Rq_{l-1})$, being
of rapid decrease in either set of variables. As a
consequence, for $R\to\infty$ and at least one $p_j$ being different
from zero, the expression approaches zero faster than any inverse
power (together with all its derivatives).

From this we see that, as $R\to\infty$, the region of possible singular
behavior is located around $(p)_1^{l-1}=0$ or $(q)_1^{l-1}=0$,
implying also $p_l=-\sum_1^{l-1}p_j=0$. We can hence infer that only
the singular behavior of $\hat{W}^T$ in $(q)=0$ will matter in this
limit. As a consequence, it will be our strategy to isolate this
singular contribution in $\hat{W}^T$ and transform it in a certain
explicit scaling behavior in $R$, which can be encoded in some power,
$R^{-\alpha}$, in front of the integral.

The singular behavior of $\hat{W}^T(q)$ at $(q)=0$ is related to the
weak decay of $W^T(y)$ at infinity. The limiting behavior of $W^T(y)$
can, however, not expected to be simple or uniform (at least not in
the generic case) as $(y_1,\ldots,y_{l-1})$ or $(x_1,\ldots,x_l)$ can
move to infinity in many different ways. We may, for example, have that
$(x_i)$ together with all $|x_i-x_j|$ go to infinity or, on the
other side, the variables move to infinity in certain fixed clusters
of finite diameter. The rate of decay of $W^T(y)$ should of course
depend in general on these details. Correspondingly, the singular
behavior of $\hat{W}^T(q)$ in the infinitesimal
neighborhood of $(q)=0$ should depend on the direction in which
$(q)=0$ is approached, that is, the limit may be direction-dependent.

In the light of this general situation we must at first decide, in
which kind of limit we are mainly interested. Inspecting the
expression (\ref{cor}), we actually started from, we choose in a
first step our fixed vectors, $(X_i)$, so that
\begin{equation}X_i-X_j\neq 0\;\text{for all}\;i,j\end{equation}
As a consequence, all distances, $|RX_i-RX_j|$, go to infinity for
$R\to\infty$. As in the preceding section, we can choose the support of
$f$ so small that, with $x_i,x_j\in supp(f)$, we have
\begin{equation}|R(X_i-x_i)-R(X_j-x_j)|\to\infty\end{equation}
In this particular case we may expect a relatively uniform limit
behavior on physical grounds.\\[0.3cm]
Remark: Similar problems occur in quantum mechanical scattering
theory.\vspace{0.3cm}

Under this proviso the following assumption seems to be reasonable.
\begin{assumption}Under the assumption, being made above, we assume
  the following decomposition of $W_l^T(y)$ to be valid: It exists a
  function, $(1+H(y))$, $H(y)$ homogeneous and positive for $y\neq 0$ so
  that
\begin{equation}G(y):=(1+H(y))\cdot W^T(y)=const+F(y)\end{equation}
with $F$ sufficiently decaying at infinity in the channel, indicated
above, i.e. $\{|y_i|\to\infty\;\text{for all}\;i=1,\ldots,l-1\}$ and
\begin{equation}H(Ry)=R^{\alpha'_l}\cdot H(y)\end{equation}
\end{assumption}
\begin{bem}A typical example for $H(y)$ is $\left(\sum
    y_i^2\right)^{\alpha'_l/2}$.
\end{bem}

Fourier transforming $G(y)$, we get
\begin{equation}\hat{G}(q)=const\cdot\delta(q)+\hat{F}(q)\end{equation}
and expression (\ref{fourcor}) becomes (compare the related expression
in \tit{Method One} of the preceding section)
\begin{multline}const\cdot
  R^{l(n-\gamma)}\cdot\int\,\prod_1^{l-1}dp_j\,\hat{G}(q)\cdot
e^{-i\sum q_jY_j}\cdot\\
\left[\int e^{-i\sum_1^{l-1} Rp_jx_j}\cdot e^{iRq_{l-1}x_l}\cdot
  (1+H(Ry+RY))^{-1}\cdot\prod_1^lf(x_j)\cdot\prod_1^ldx_j\right]\end{multline}

By assumption, $H$ is homogeneous of degree $\alpha'_l$. So we can
extract a negative power of $R$, $R^{-\alpha'_l}$, from the expression
in square brackets. Furthermore, we observed above that for
$R\to\infty$ only the vicinity of $q=0$ matters. Finally, by
assumption, the contribution coming from $\hat{F}(q)$ can be neglected
in this limit (compare the corresponding discussion in the subsection
\ref{One}; as a consequence of the assumed support properties, the
expression in square brackets is again strongly decreasing). We hence have
\begin{multline}\lim_{R\to\infty}\langle A_R(R\cdot X_1)\cdots A_R(R\cdot
  X_l)\rangle^T=\lim_{R\to\infty}const\cdot
  R^{(ln-\alpha'_l-l\gamma)}\cdot
  \int\prod_1^{l-1}dq_j\\\delta(q)\cdot  e^{-i\sum_{j=1}^{l-1}Rq_jY_j}\cdot\left[\int
    e^{-i\sum_1^{l-1}Rp_jx_j}\cdot
    e^{iRq_{l-1}x_l}\cdot\left(R^{-\alpha'_l}+H(y+Y)\right)^{-1}\cdot \prod_{i=1}^{i=l}
    f(x_i)\prod_{i=1}^{i=l}dx_i\right]\end{multline}
Remark: We see again the reason for the special choice being made
above as to the support properties of the functions $f(x_i)$, leading to the
result $y_j+Y_j\neq 0$ on the support of $f$. Without this assumption,
we see for our above example, $H(y)=\left(\sum
    y_i^2\right)^{\alpha'_l/2}$, that in the limit, where
  $R^{-\alpha'_l}$ vanishes, we would get a singular contribution at
  points where $y+Y=0$ in the integrand in square brackets. These terms
  would make the following discussion much more tedious.\vspace{0.3cm}

If we now make the choice
\begin{equation}\gamma:=\gamma_l=n-\alpha'_l/l\end{equation}
we arrive at a finite limit expression, depending on the coordinates $(X_i)$:
\begin{equation}\lim_{R\to\infty}\langle A_R(R\cdot X_1)\cdots A_R(R\cdot
  X_l)\rangle^T=const\cdot\int
  H_l(y+Y)^{-1}\cdot\prod_1^lf(x_i)\prod_1^ldx_i\end{equation}
which is a function of the coarse grained difference coordinates
\begin{equation}Y_j=X_j-X_{j+1}\end{equation}
For the $Y_j$ sufficiently large, it is approximately a function
\begin{equation}W_{limit}(Y)\approx const\cdot
  H_l(Y)^{-1}\end{equation}
That is, the renormalized limit correlation functions reproduce the
asymptotic power law behavior of the original microscopic correlation
functions.

For later use we introduce the new scaling exponent, $\alpha_l$, via
\begin{equation}\alpha'_l+\alpha_l=(l-1)n\end{equation}
This implies
\begin{equation}\label{gamma}\gamma_l=(n+\alpha_l)/l\end{equation}
The underlying reason for this choice is that an asymptotic decay,
$\sim r^{-(l-1)n}$, is just the threshold for $W_l^T$ being integrable
or non-integrable (with $r:=\left(\sum y_j^2\right)^{1/2}$).

We have arrived at the following result: We are interested in a
scaling-limit theory for $R\to\infty$. In order to get a non-vanishing
and finite limit theory, we have to choose the scaling exponent for $l=2$ as
\begin{equation}\gamma=\gamma_2=(n+\alpha_2)/2\end{equation}
Furthermore, we have extracted the asymptotic form from the higher
truncated $l$-point functions, $W_l^T(y)$, and have absorbed it in an
explicit scaling factor, $R$ to some power. If the limit theory is to
be finite, the corresponding scaling exponents for $l>2$ have to be
less or equal to zero. This yields unique $\gamma_l$'s as threshold
values.

A corner stone of the philosophy of the renormalisation group is that
the scaling exponents of the scaled observables remain the same,
irrespectively of the degree of the correlation functions in which
they occur. That is, these exponents are fixed by the exponent,
$\gamma_2$, and we have
\begin{equation}\gamma=\gamma_2\geq \gamma_l\end{equation}
(the latter exponent being derived from equation (\ref{gamma})),
in order that the limit correlation functions remain finite.
\begin{conclusion}We have the following alternatives for $R\to\infty$:
\begin{align}\gamma_2>\gamma_l & \Rightarrow & W_{l,R}^T\to
  0\\
\gamma_2=\gamma_l & \Rightarrow & \lim_{R\to\infty}
  W_{l,R}^T\quad\text{is finite and non-trivial}\\
\gamma_2<\gamma_l & \Rightarrow &  W_{l,R}^T\to\infty\end{align}
If $\gamma_2>\gamma_l$ for all $l\geq 3$, the fixed point is gaussian or
trivial. The limit theory is quasi-free. The limit theory is
non-trivial if $\gamma_2=\gamma_l$ for at least some $l\geq 3$. For
$\gamma_2<\gamma_l$ for some $l$, the limit theory does not exist.
\end{conclusion}
\begin{bem}The corresponding analysis can also be done by employing
  \tit{Method Two} (discussed in the preceding section). One can even
  omit the support conditions assumed above. The treatment then
  becomes more involved but the end result is the same. We discuss one
  particular case below.
\end{bem}

To complete the scaling and/or cluster analysis of the truncated
correlation functions, we have to analyze the other channels and the
respective consequences for scaling exponents and cluster
assumptions.

We mentioned several times that without the support condition
\begin{equation}(X_i-X_j)+(x_i-x_j)\neq 0 \end{equation}
for $x_{i,j}\in supp(f)$, the analysis would become more tedious. On
the other side, this assumption is violated if the observables move to
spatial infinity in certain clusters. The extreme case occurs when all
$X_i$ are chosen to be zero, i.e:
\begin{equation}\langle A_R(1)\cdots
  A_R(l)\rangle^T\;,\;R\to\infty\end{equation}
(the indices 1,\ldots,l denote the different observables).
This scenario was already briefly discussed in section 7 of \cite{Re1}
in connection with phase transitions and/or spontaneous symmetry
breaking, which are also typically related to poor spatial clustering.

With the same notations as above we have
\begin{multline}\langle A_R(1)\cdots A_R(l)\rangle^T=const\cdot R^{l(n-\gamma)}\cdot\\\int
  \hat{W}_l^T(q_1,\ldots,q_{l-1})\cdot
  \left[\int
    e^{-i\sum_1^{l-1}Rp_jx_j}\cdot e^{iRq_{l-1}x_l}\cdot
    \prod_{i=1}^{i=l}
    f(x_i)\prod_{i=1}^{i=l}dx_i\right]\prod_1^{l-1}dp_j\end{multline}
Assuming again the existence of a suitable homogeneous function,
$H_l(y)$, in this channel, we get asymptotically two contributions
\begin{equation}\label{term1}const\cdot R^{l(n-\gamma)-\alpha'_l}\cdot\int
 H_l(y)^{-1}\cdot\prod_1^lf(x_i)\prod_1^ldx_i\end{equation}
and
\begin{multline}\label{term2}const\cdot
  R^{l(n-\gamma)-\alpha'_l}\cdot\int\prod_1^{l-1}dq_j\,
  \hat{F}(q_1,\ldots,q_{l-1})\cdot\\
\left[\int
    e^{-i\sum_1^{l-1}Rp_jx_j}\cdot
    e^{iRq_{l-1}x_l}\cdot\left(H_l(y)\right)^{-1}\cdot \prod_{i=1}^{i=l}
    f(x_i)\prod_{i=1}^{i=l}dx_i\right]\end{multline}

The first term has almost the same form as above. But now the function
in square brackets in the second contribution is no longer of strong
decrease as the integrand (considered as a function of $(x)$or $(y)$)
is no longer in $\mcal{D}$ as it will have a singularity in $y=0$. We
can however provide the following estimate on the degree of this
singularity of $H_l^{-1}$ in $y=0$. We assumed throughout in this
section that $W_l^T$ is not integrable at infinity, that is, the
clustering is weak. On the other side, this asymptotic behavior is
exactly encoded in $H_l^{-1}$, as we observed above. The threshold where
integrability goes over into non-integrability for $H^{-1}_l$ is a
behavior
\begin{equation}\sim
  r^{-(l-1)n}\;,\;r:=\left(\sum_1^{l-1}y_j^2\right)1/2\end{equation}

We can therefore conclude that
\begin{equation}\alpha'_l\leq (l-1)n\end{equation}
in the above construction if $W_l^T$ is non-integrable at infinity. If
$\alpha'_l$ is even strictly smaller than $(l-1)n$, which is the ordinary
case in the critical region, we have
\begin{ob}
\begin{equation}\alpha'_l< (l-1)n\end{equation}
implies that $H_l^{-1}$ is integrable near $y=0$. Hence
\begin{equation}H_l^{-1}(y)\cdot \prod_1^l f(x_i)\in L^1\end{equation}
due to the compact support of $f$.
\end{ob}

From this we infer again that, with
\begin{equation}\gamma_l=n-\alpha'_l/l=(n+\alpha_L)/l\end{equation}
the contribution (\ref{term1}) is finite in the scaling limit. For the
contribution (\ref{term2}) we have by the same reasoning that the
function in square brackets is a continuous function of $(Rq)$, which
goes to zero for $Rq\to\infty$ (due to the Riemann-Lebesgue lemma).

On the other side, we have no precise apriori information about $F(y)$
and $\hat{F}(q)$. $F(y)$ goes to zero at infinity as the asymptotic
behavior is contained in $H^{-1}$, but its rate of vanishing is not
clear.
\begin{conclusion}If the integrand of contribution (\ref{term2}) is
  lying in some $L^p$, so that the limit, $R\to\infty$, can be
  performed under the integral, the whole expression vanishes in the
  scaling limit.
\end{conclusion}
In this situation we are left with again with the first term, which
is the limit of
\begin{equation}const\cdot\int
  H_l(y+Y)^{-1}\cdot\prod_1^lf(x_i)\prod_1^ldx_i\end{equation}
for $Y\to 0$. That is, in this case it holds
\begin{satz}If the situation is as in the conclusion,
  $W_l^{lim}(X_1,\ldots,X_l)$ is continuous and we have in particular
\begin{equation}W_l^{lim}(0,\ldots,0)=\lim_{X\to
    0}W_l^{lim}(X_1,\ldots,X_l)\end{equation}
\end{satz}

We can hence resume our findings as follows: If the assumptions, made
above, are fulfilled and if the functions, $H_l$, can be chosen
consistently in all channels, so that the $\gamma_l$'s, resulting from the
relation
\begin{equation}\gamma_l=(n+\alpha_l)/l\end{equation}
are smaller than or identical to $\gamma_2$, we arrive at a full limit
theory, being well-defined in all channels. In this case the
renormalization group program works and yields a non-trivial scaling
limit.

\end{document}